\documentclass[5p]{elsarticle}
\usepackage{amsmath,subfloat,enumitem,hyperref}
\usepackage[numbers]{natbib}
\usepackage[caption=false]{subfig}

\begin{document}
\begin{frontmatter}
\title{Origin of chaos in 3-d Bohmian trajectories}

\author{Athanasios C. Tzemos}
\ead{thanasistzemos@gmail.com}
\author{George Contopoulos}
\ead{gcontop@academyofathens.gr}
\author{Christos Efthymiopoulos}
\ead{cefthim@academyofathens.gr}
\cortext[cor1]{A.C. Tzemos , Tel: +306932271598}
\address{Research Center for Astronomy and Applied Mathematics of the Academy of Athens, Soranou Efesiou 4, GR-11527 Athens, Greece}

\begin{abstract}
We study  the 3-d Bohmian trajectories of a quantum system of three harmonic oscillators. We focus  on the mechanism responsible for the generation of chaotic trajectories. We demonstrate the existence of a 3-d analogue of the mechanism found in earlier studies of 2-d systems \cite{efthymiopoulos2007nodal, PhysRevE.79.036203}, based on moving 2-d `nodal point - X-point complexes'. In the 3-d case, we observe a foliation of nodal point - X-point complexes, forming a `3-d structure of nodal and X-points'. Chaos is generated when the Bohmian trajectories are scattered at one or more close encounters with such a structure. 
\end{abstract}

\end{frontmatter}

\section{\label{sec:level1}Introduction}

The Bohmian picture of quantum mechanics \cite{Bohm,BohmII} has attracted the attention of many researchers in the last two decades, due not only to its potential theoretical interest \cite{holland1995quantum, deurr2009bohmian}, but also to its wide spectrum of applications in the fields of atomic and molecular physics, nanoelectronics, light-matter interaction etc (see \cite{benseny, pladevall2012applied}). In the Bohmian picture, besides the evolution of the particle wavefunction, governed by the Schr\"{o}dinger equation
\begin{align}\label{sch}
\Big[-\frac{\hbar^2}{2m}\nabla^2+V\Big]\Psi(\vec{x},t)=i\hbar\frac{\partial \Psi(\vec{x},t)}{\partial t},
\end{align}
we consider also particle trajectories governed by the `guidance' equation 
\begin{align}\label{gui}
m\frac{d\vec{x}}{dt}=\hbar\Im\Big(\frac{\nabla\Psi(\vec{x},t)}{\Psi(\vec{x},t)}\Big).
\end{align}
In the Bohmian interpretation, these trajectories acquire `ontological' significance as part of the physical reality of microscopic systems. However, the Bohmian trajectories can also be viewed as tracers of the  quantum probability flow, a fact allowing to regard the Bohmian picture as a trajectory-based approach for studying the time evolution of quantum processes also in the framework of the conventional interpretation of quantum mechanics (see \cite{sanz2012trajectory} ).

An important topic in Bohmian mechanics is the emergence of {\it chaos} and the role played by chaotic Bohmian trajectories. It has been suggested that Bohmian chaos is relevant in the interpretation of a variety of quantum phemomena. Examples are: i) the precision limits of hydrodynamical solvers of Schr\"{o}dinger's equation (e.g. \cite{pladevall2012applied, trahan2005quantum, oriols2007quantum}), and ii) `quantum relaxation', i.e., the possibility to assign a dynamical origin to Born's rule for the quantum probabilities \cite{valentini2005dynamical}. More examples are dicussed in \cite{benseny}.  

The mechanism of generation of chaos is thought to be well understood in 2-d quantum systems \cite{efthymiopoulos2007nodal,PhysRevE.79.036203,durr1992quantum,wu1994inverse,
iacomelli1996regular, frisk1997properties, wisniacki2005motion, efthymiopoulos2006chaos, wisniacki2007vortex, contopoulos2008ordered,    contopoulos2012order}. However, only few works exist in the 3-d or higher dimensional cases, due to the far higher complexity of the problem \cite{frisk1997properties, bialynicki2000motion, falsaperla2003motion, 2016arXiv160301387C}.

In this letter we study an example of  3-d quantum system, aiming to investigate the origin of chaos and the dynamical properties of 3-d chaotic Bohmian trajectories.

The Hamiltonian consists of three harmonic oscillators
\begin{align}\label{ham}
H=\sum_{k=1}^3\Big(\frac{p_k^2}{2m}+\frac{1}{2}m_k\omega_k^2x_k^2\Big).
\end{align}
As exemplified in section 2 below, the quantum states of the system  (\ref{ham}) typically exhibit both regular and chaotic Bohmian trajectories. Focusing on the latter, in our study below we extend the results found in corresponding 
2-d cases, which in summary are the following:

1. In the vicinity of the nodal points (i.e. the points in configuration space where the  wavefunction becomes null $\Psi=0$) we observe, in general, the formation of quantum vortices. A vortex around a nodal point is visualized by drawing the 2-d vector field of the probability flow in a system of reference co-moving with the nodal point. 

2. Near every  nodal point, we can prove the existence of a second critical 
point of the flow in the comoving frame. This second point is always hyperbolic, and it was called the X-point in refs. \cite{efthymiopoulos2007nodal,PhysRevE.79.036203}. The chaotic behavior of the trajectories is due to the cumulative effect of their repeated close encounters with the X-points (while, as noted already in \cite{Bohm, BohmII}, the trajectories avoid, in general, coming close to the corresponding moving nodal points). In particular, the `stretching number' (i.e. the local Lyapunov exponent) of a trajectory undegoes a positive kick at every such close encounter. The above mechanism of chaos is generic for 2-d systems, in the sense that its existence is guaranteed by a local analysis of the quantum  flow near every nodal point generated by an arbitrary 2-d wavefunction \cite{PhysRevE.79.036203}. 

3. The 2-d structure formed by a nodal point and its associated X-point, as well as the latter's stable and unstable manifolds and all nearby streamlines of the quantum probability flow,  is called a `nodal point - X-point complex'. A demonstrated in \cite{PhysRevE.79.036203}, the nodal point-X-point structure of the quantum flow is generic close to \textit{moving} nodal points, i.e. this structure  necessarily appears in the frame of reference comoving with any nodal point. On the other hand, we may have different topologies appearing close to fixed nodal points, which however, cannot lead to chaos.

In the present letter, we show that these results can be extended to the 3-d case. 

An important early work in the 3-d case was provided by Falsaperla and Fonte \cite{falsaperla2003motion}, who made a careful analysis of the quantum flow, based on a leading order expansion  of the Bohmian equations of motion  around nodal points. These authors found that: i) the streamlines of the quantum flow lie locally on planes orthogonal to a ``nodal line'', i.e. a line formed in the 3-d configuration space by continuously joining  the nodal points  of the solutions of $\Psi=0$ for a given time. ii) The local form of the streamlines is  spiral and the union of all these spirals surrounds the nodal line forming a ``cylindrical structure''. This structure influences the form of the trajectories close to nodal lines. However, according to \cite{falsaperla2003motion}, the dynamical behavior of the trajectories close to the nodal points appears ``regular'', while chaos is attributed to an ``intermittent'' ``switch back and forth'' from the vicinity of one nodal line to that of another nodal line.   

Here we demonstrate that, similarly to the 2-d case, in each of the orthogonal planes discussed already in \cite{falsaperla2003motion}, the structure of the quantum flow is such as to  form, besides the nodal point, also an X-point, and hence, a complete 2-d nodal point X-point complex. We compute numerically several such complexes for various times and in various locations of the 3-d configuration space and establish that their topological structure is similar to those established in the 2-d case. However, in the 3-d case, the ensemble of all the orthogonal planes forms a foliation. Then, besides joining the nodal points along this foliation, thus forming a `nodal line', we can  also join the X-points, thus forming an `X-line'. The whole geometrical structure of the quantum flow formed in this way is hereafter called a `3-d structure of nodal and X-points'. This structure extends the ``cylindrical structure'' observed in \cite{falsaperla2003motion}, by adding the X-points to it. This extension is crucial for understanding the generation of chaos. Namely, chaos is produced by the repeated scatterings of the Bohmian trajectories with the X-points along the 3-d structure of nodal and X-points. In fact, this mechanism implies that chaotic trajectories in 3-d systems can exist even when only  one such structure is present for every time, i.e., one  does not need the intermittent jumps between more than one such structure, that were conjectured in \cite{falsaperla2003motion}. On the other hand, we leave open the question of how general our presently proposed chaos mechanism is.

In section 2, we consider a basic 3-d quantum state of the Hamiltonian~(\ref{ham}) in which the orthogonal relation between  the quantum flow and the nodal lines is exact. This model serves to establish the basic features of the chaos mechanism and to show the  `3-d structure of nodal and X-points'. Yet, this model is not generic since Bohmian trajectories cannot exhibit any motion in the direction normal to the orthogonal planes, i.e. along the 3-d structure of nodal and X-points. However, in section 3 we perturb our previous state in a way so as to secure the occurrence of genuinely 3-d chaotic diffusion. Finally, section 4 summarizes the conclusions of our present study.

\section{Basic Model}

\textit{Equations of motion:} In the Hamiltonian (\ref{ham}) we studied the Bohmian trajectories in several cases of wavefunctions of the form
 \begin{align}\label{wavefunction}
\Psi(\vec{x},t)&=\sum_{i=1}^3a_ie^{-\frac{i}{\hbar}E_it}\Psi_{i}(\vec{x}),\quad \sum_{i=1}^3|a_i|^2=1
 \end{align}
 where  $\Psi_i(\vec{x}) ,i=1..3$  are eigenstates of the form
\begin{align}\label{eigenstate}
\Psi(\vec{x})=\prod_{k=1}^3\frac{\Big(\frac{m_k\omega_k}{\hbar\pi}\Big)^{\frac{1}{4}}\exp\Big(\frac{-m_k\omega_kx_k^2}{2\hbar}\Big)}{\sqrt{2^{n_k}n_k!}}H_{n_k}\Big(\sqrt{\frac{m_k\omega_k}\hbar{}}x_k\Big),
\end{align} 
where $m_k$ are the masses, $\omega_k$ the frequencies of the oscillators and $H_{n_k}$ are the Hermite polynomials of order $n_k$. The energy of the eigenstate (\ref{eigenstate}) is $E=\sum_{k=1}^3(n_k+\frac{1}{2})\hbar\omega_k.$
We  work in units in which $m_1=m_2=m_3=\hbar=1$, while we set $\omega_1=1, \omega_2=\sqrt{2}, \omega_3=\sqrt{3}$. 

We computed trajectories in several examples of wavefunctions of the form (\ref{wavefunction}), and we typically found both ordered and chaotic Bohmian trajectories. In the sequel, we focus on one particular of these examples, namely

\begin{align}\label{state}
\Psi(\vec{x},t)=\frac{\Psi_{1,0,0}(\vec{x},t)+\Psi_{0,1,0}(\vec{x},t)+\Psi_{0,0,1}(\vec{x},t)}{\sqrt{3}}.
\end{align}

The wavefunction (\ref{state}) turns to be convenient in our analysis below due to the following property: The Bohmian equations of motion, derived from Eq.(\ref{gui}) are:

\begin{align}\label{troxia1}
&\dot{x_i}=\frac{1}{G}\sum_{\substack{j=1\\ i\neq j}}^3\sqrt{(\omega_j\omega_i)}\sin(\omega_{ji}t)x_j, 
\end{align} where $G=\sum_{\substack{i=1}}^3\Bigg(\sum_{\substack{j=1\\ j\neq i}}^3\Big(\sqrt{\omega_i\omega_j}\cos(\omega_{ij}t)x_ix_j+\omega_i^2x_i^2\Big)\Bigg)$ and  $\omega_{ij}=\omega_i-\omega_j$.

From Eqs.~(\ref{troxia1}) we derive that $\sum_{i=1}^3x_i\dot{x}_i=0$. 
Therefore all the Bohmian trajectories lie on  spherical surfaces \begin{align}\label{sphere1}
\sum_{i=1}^3x_i^2=R^2
\end{align}
with $R$ determined from the initial conditions.

\textit{ Nodal point-X-point complexes:}
The nodal points are the solutions, for $(x_1, x_2, x_3)$, of the system of equations 
\begin{align}\label{nodalsystem}
Re(\Psi(x_1, x_2, x_3,t))=0, \quad
Im(\Psi(x_1, x_2, x_3,t))=0.
\end{align}
Since  (\ref{nodalsystem}) is a system of 2 equations for 3 variables, the solutions lie, in general, along curves in the 3-d space. In our particular case any solution $x_1=x_{1nod}, x_2=x_{2nod}, x_3=x_{3nod}$ of (\ref{nodalsystem}) satisfies the relation
\begin{align}
\frac{\sqrt{\omega_1}x_{1nod}}{\sin(\omega_{32}t)}&=\frac{\sqrt{\omega_2}x_{2nod}}{\sin(\omega_{13}t)}=\frac{\sqrt{\omega_3}x_{3nod}}{\sin(\omega_{21}t)}=C\label{straight}.
\end{align}
Consequently the nodal points all lie along straight lines in 3-d space passing through the origin $(0,0,0)$, whose inclination changes with the time  $t$.

\begin{figure}[hbt]\centering
\captionsetup[subfloat]{captionskip=-2.4cm}
\subfloat[]{\label{fig:troxia_nodal_point}\includegraphics[scale=0.1689]{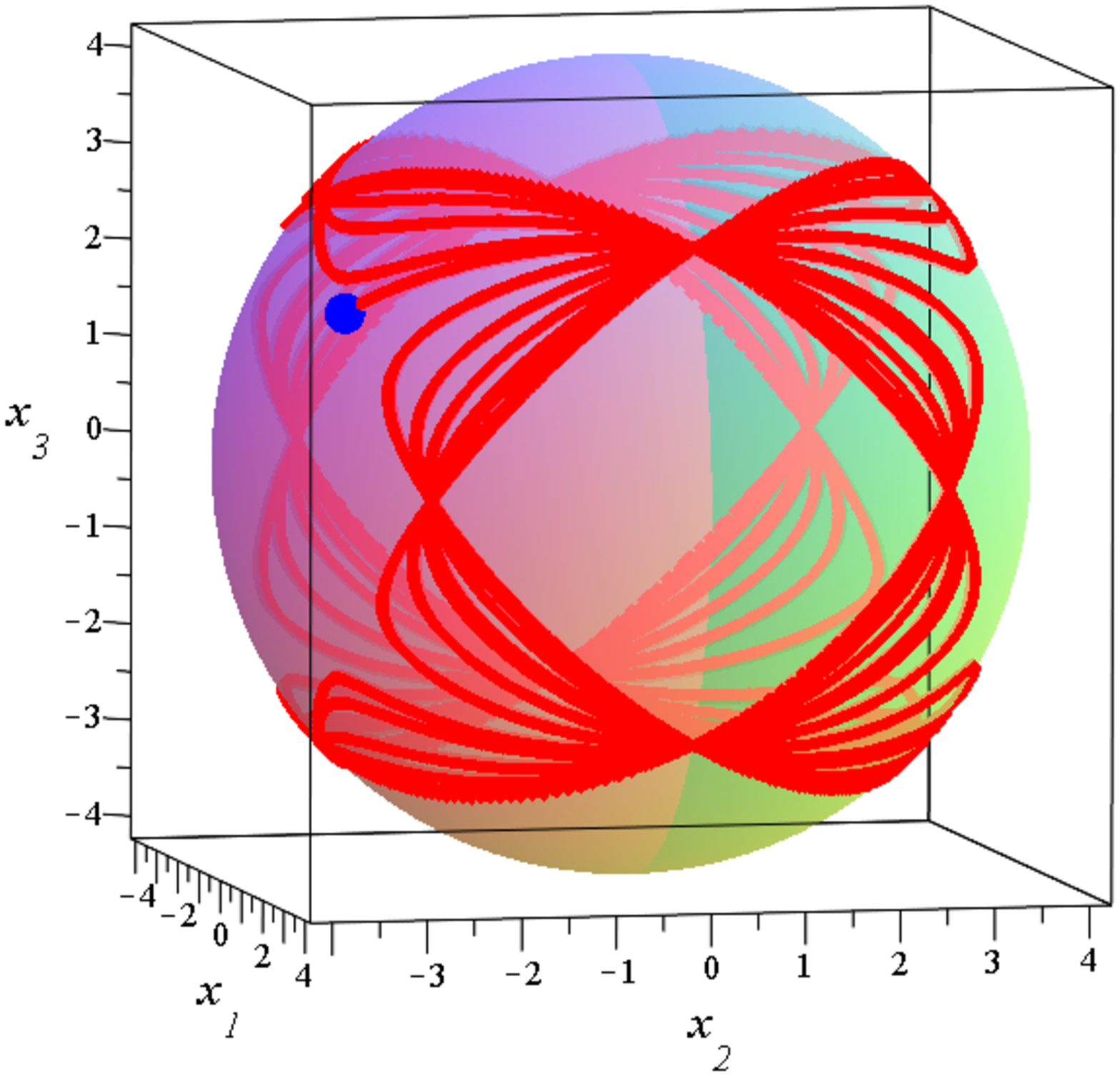}}\hfill
\subfloat[]{\label{fig:acceleration}\includegraphics[scale=0.1999]{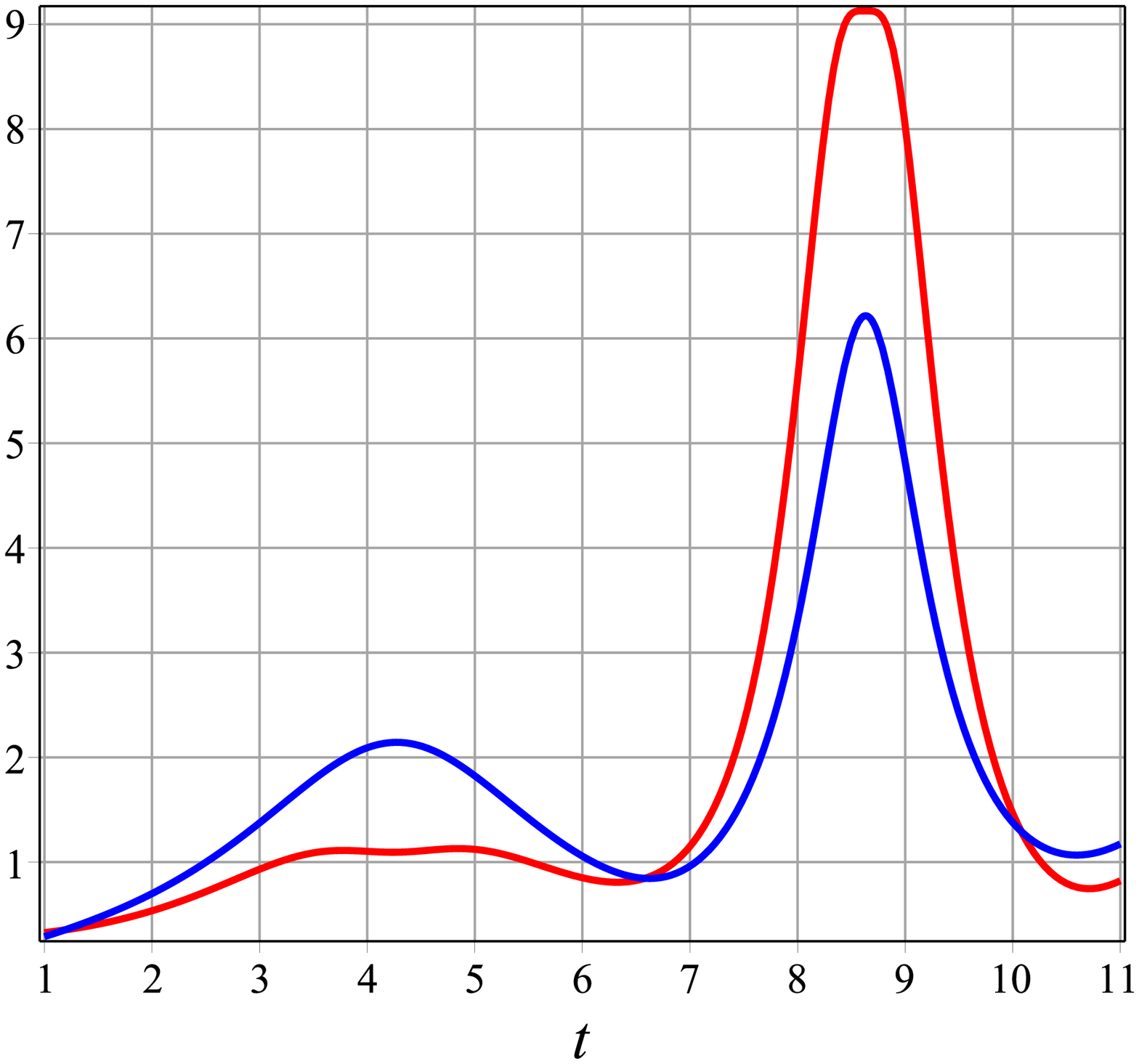}}
\caption{(a)  The trajectory of a nodal point starting at a distance 4.23 from (0,0,0) for $t=1$ (blue point) up to $t=250$. This trajectory is on the surface of a sphere (strong red lines on the foreground  and weaker lines on the background  of the sphere). (b) The velocity (blue curve) and the acceleration (red curve) of a nodal point  as  functions of time. Both increase abruptly near a convergence point at $t\simeq 8.5$.}
\end{figure}

Consider, now, the intersection of such a line with the surface of the sphere of fixed radius $R$. The point of intersection is a nodal point describing a trajectory on the sphere. Due to Eq.(\ref{straight}) we have
$\sqrt{\omega_2\omega_3}\sin{(\omega_{32}t)}\dot{x}_{1nod}+\\\sqrt{\omega_1\omega_3}\sin{(\omega_{13}t)}\dot{x}_{2nod}+\sqrt{\omega_1\omega_2}\sin{(\omega_{21}t)}\dot{x}_{3nod}=0.$
Thus, the velocity of the nodal point is perpendicular to the corresponding nodal line.
Figure \ref{fig:troxia_nodal_point} shows  the trajectory of the nodal point on the surface $R=4.23$ from $t=1$ (blue point) up to $t=250$. This trajectory passes several times through some convergence points,  near which the velocity and acceleration of the nodal point increases considerably (Fig.~\ref{fig:acceleration}).

We will now examine the structure of the quantum flow in a moving plane tangent to the sphere of fixed radius $R$ (like the one of Fig.~\ref{fig:troxia_nodal_point}), at a point coinciding with the instantaneous position of the nodal point on the same sphere at any time $t$ (this plane coincides with the plane `orthogonal to the nodal line' defined in \cite{falsaperla2003motion}). To this end, we define a non-inertial frame  of reference $(x_1', x_2', x_3')$ whose $x_3'$-axis  coincides, at any time $t$, with the  corresponding nodal line of Eq.~(\ref{straight}).
The change of coordinates $(x_1,x_2,x_3)\to (x_1',x_2',x_3')$ is realized via the transformation
\begin{align}\label{transf}
&\vec{x'}=S\vec{x},\quad
\dot{\vec{x}'}=S\dot{\vec{x}},
\end{align}
where 
$
S(t)=\Bigg(\begin{smallmatrix}
    \sin{\phi(t)} & -\cos{\phi(t)} & 0   \\
    \cos{\theta(t)}\cos{\phi(t)} & \cos{\theta(t)}\sin{\phi(t)} & -\sin{\theta(t)} \\
    \sin{\theta(t)}\cos{\phi(t)} & \sin{\theta(t)}\sin{\phi(t)} & \cos{\theta(t)}
\end{smallmatrix}\Bigg),
$
with $\cos\theta(t)=\frac{x_{3nod}(t;R)}{\sqrt{\sum_{i=1}^3x_{inod}(t;R)^2}}$ and $\tan\phi(t)=\frac{x_{2nod}(t;R)}{x_{1nod}(t;R)}$, $ \theta\in[0,\pi]$ and $\phi\in[0,2\pi)$. Setting $\vec{U}=(u,v,w)$, where $u=x_1-x_{1nod}(t), v=x_2-x_{2nod}(t), w=x_3-x_{3nod}(t)$, 
$A(t)=\Bigg(\begin{smallmatrix}
0 & -\sqrt{\omega_1\omega_2}\sin(\omega_{12}t) & -\sqrt{\omega_1\omega_3}\sin(\omega_{13}t)\\
\sqrt{\omega_1\omega_2}\sin(\omega_{12}t) & 0 & -\sqrt{\omega_2\omega_3}\sin(\omega_{23}t)\\
\sqrt{\omega_1\omega_3}\sin(\omega_{13}t) & \sqrt{\omega_2\omega_3}\sin(\omega_{23}t) & 0
\end{smallmatrix}\Bigg)$ and using (\ref{straight}), Eqs.~(\ref{troxia1}) take the form
\begin{align}
\dot{\vec{U}}'&=S(t)\dot{\vec{U}}=\frac{1}{G}S(t)A(t)S^{-1}(t)\vec{U}'-S(t)\dot{\vec{x}}_{nod}.
\end{align}
Consequently
\begin{align}\label{equationtr}
&\Bigg(\begin{smallmatrix}
\dot{u'}\\\dot{v'}\\\dot{w'}
\end{smallmatrix}\Bigg)=\nonumber
\Bigg(\begin{smallmatrix}
f_1(u',v',w',t)\\
f_2(u',v',w',t)\\
f_3(u',v',w',t)
\end{smallmatrix}\Bigg)\\&=\frac{1}{G}S(t)A(t)S^{-1}(t)\Bigg(\begin{smallmatrix}
u'\\
v'\\
w'
\end{smallmatrix}\Bigg)-S(t)\Bigg(\begin{smallmatrix}
\dot{x}_{1nod}(t;R)\\
\dot{x}_{2nod}(t;R)\\
\dot{x}_{3nod}(t;R)
\end{smallmatrix}\Bigg)
\end{align}
and $G$ takes the form $G=u'^2\Phi_1+u'v'\Phi_2+v'^2\Phi_3$
with
$
\Phi_1=-\sqrt{\omega_1\omega_2}\cos(\omega_{12}t)\sin2\phi(t)+\omega_1\sin^2\phi(t)\\+\omega_2\cos^2\phi(t),\label{Phi1}\\
\Phi_2=\nonumber2\sqrt{\omega_2\omega_3}\cos(\omega_{23}t)\cos\phi(t)\sin\theta(t)\\-2\sqrt{\omega_1\omega_3}\cos(\omega_{13}t)\sin\phi(t)\sin\theta(t)\\+2\sqrt{\omega_1\omega_2}\cos(\omega_{12}t)(-\cos\theta(t)\cos^2\phi(t)\\+\cos\theta(t)\sin^2\phi(t))+\cos\theta(t)\sin2\phi(t)(\omega_1-\omega_2),\label{Phi2}\\
\Phi_3=-\sqrt{\omega_2\omega_3}\cos(\omega_{23}t)\sin2\theta(t)\sin\phi(t)\\-\sqrt{\omega_1\omega_3}\cos(\omega_{13}t)\sin2\theta(t)\cos\phi(t)\\+\sqrt{\omega_1\omega_2}\cos(\omega_{12}t)\cos^2\theta(t)\sin2\phi(t)\\ \nonumber+\omega_1\cos^2\theta\cos^2\phi(t)+\omega_2\cos^2\theta(t)\sin^2\phi(t)+\omega_3\sin^2\theta(t).\label{Phi3}
$

The equations (\ref{equationtr}) are subject to the constrain (\ref{sphere1}) expressed in the new variables $(u',v',w')$. Thus, only two out of these three variables are independent. Also, up to terms of second order in $u'/R$ or $v'/R$, the constrain (\ref{sphere1}) yields $w'=\dot{w'}\simeq0$.  Then the first and second equations of the system (\ref{equationtr}) approximately become:

\begin{align}\label{reducedsystem}
\Big(\begin{smallmatrix}
\dot{u'}\\\dot{v'}\end{smallmatrix}\Big)\simeq\Big(\begin{smallmatrix}
F_1(u',v',t)\\F_2(u',v',t)
\end{smallmatrix}\Big)=\frac{1}{G}\Big(\begin{smallmatrix}
0& B(t)\\-B(t)& 0
\end{smallmatrix}\Big)\Big(\begin{smallmatrix}
u'\\v'
\end{smallmatrix}\Big)-\Big(\begin{smallmatrix}V_u(t)\\V_v(t)\end{smallmatrix}\Big)
\end{align}
where
$ B(t)=-\cos\theta(t)\sqrt{\omega_1\omega_2}\sin(\omega_{12}t)\\\nonumber+\sin\theta(t)\sin\phi(t)\sqrt{\omega_1\omega_3}\sin(\omega_{13}t)\\-\sin\theta(t)\cos\phi(t)\sqrt{\omega_2\omega_3}\sin(\omega_{23}t),$ 
\begin{align}
&V_u=\nonumber\dot{x}_{1nod}'=\sin\phi(t)\dot{x}_{1nod}-\cos\phi(t)\dot{x}_{2nod}\\
&V_v=\nonumber\dot{x}_{2nod}'=\cos\theta(t)\cos\phi(t)\dot{x}_{1nod}+\cos\theta(t)\sin\phi(t)\dot{x}_{2nod}\\&\nonumber-\sin\theta(t)\dot{x}_{3nod}
\end{align}
and $\dot{x}_{1nod}, \dot{x}_{2nod}, \dot{x}_{3nod}$ are functions of time found with the help of Eqs.(\ref{straight}).

The key remark is that Eqs.(\ref{reducedsystem}) are of the same form as the equations (4) of Ref.\cite{PhysRevE.79.036203}. Thus, the analysis of the local structure of the quantum flow of Ref.~\cite{PhysRevE.79.036203} can be transferred to the present example as well. 

In particular:
near every nodal point there exists a second type of critical point of the flow, where $\dot{u'}=\dot{v'}=\dot{w'}=0$.
In the approximation of Eqs.(\ref{reducedsystem}), we have $w'=0$, so we can obtain an approximate solution $u'=u'_X, v'=v'_X$ with $F_1(u'_X,v'_X)=F_2(u'_X,v'_X)=0$, namely

\begin{align}
\label{xps}\frac{1}{G}B(t)v'_X-V_u(t)=-\frac{1}{G}B(t)u'_X-V_v(t)=0
\end{align}
From the above equations we get
$\label{vx}
v_X'=-\frac{V_u(t)}{V_v(t)}u'_X, u'_X=-\frac{B(t)}{V_v(t)\Big(\Phi_1-\frac{V_u(t)}{V_v(t)}\Phi_2+\frac{V_u(t)^2}{V_v(t)^2}\Phi_3\Big)}
$ and $w'_X=0$.
As in \cite{PhysRevE.79.036203}, we define the  Jacobian Matrix at $(u'_X,v'_X)$.
\begin{align}
J=\begin{pmatrix}
\dfrac{\partial F_1}{\partial u'}&\dfrac{\partial F_1}{\partial v'}\\
&\\
\dfrac{\partial F_2}{\partial u'}&\dfrac{\partial F_2}{\partial v'}
\end{pmatrix}\Bigg|_{(u'=u'_X,v'=v'_X)}{=}\begin{pmatrix}
a&b\\c&d\end{pmatrix}
\end{align}
The characteristic polynomial of $J$ 
\begin{align}\label{cp}
\lambda^2-(a+d)\lambda+ad-bc=0
\end{align}
yields two eigenvalues $\lambda_1$ and  $\lambda_2$ which can be proven to be always real and of opposite sign. Consequently  the `X-point' $(u'=u'_X, v'=v'_X$) is a hyperbolic point.

The denominator $G$ of Eqs. (\ref{reducedsystem}) is quadratic in terms of $u',v'$, while their numerator is linear. This implies that the velocity of a trajectory tends to infinity at small distances from the nodal point. We can then approximate locally the trajectory by a so called `adiabatic approximation'. This means to `freeze' the time in the right hand side only of Eqs. (\ref{reducedsystem}),  treating the flow as nearly autonomous  for a small time interval around a given t (see \cite{PhysRevE.79.036203} for a discussion a the error of this approximation). Then Eqs.~(\ref{reducedsystem}) take the form
\begin{align}\label{fic1}
\frac{du'}{ds}=F_1(u',v';t),\quad \frac{dv'}{ds}=F_2(u',v';t)
\end{align}
where $s$ represents now the time in a small interval around $t$, while $t$ itself is kept fixed.
With the above approximation, Eqs.~(\ref{fic1}) allow to draw phase portraits yielding the structure of the nodal point-X-point complex.
Along the eigendirections corresponding to the eigenvalues of Eq.(\ref{cp}) the (linear) deviations $\Delta u'$ and $\Delta v'$ satisfy the relation $\begin{pmatrix}
a & b\\c &d
\end{pmatrix}\begin{pmatrix}
\Delta u'\\
\Delta v'
\end{pmatrix}=\lambda \begin{pmatrix}
\Delta u'\\\Delta v'
\end{pmatrix}$.
Hence the eigenvector associated with the eigenvalue $\lambda$ has a slope $\frac{\Delta v'}{\Delta u'}=\frac{\lambda -a}{b}.$
By taking many initial values of $\Delta u', \Delta v'$ (of order $10^{-5}$), and integrating  (\ref{fic1}),  we then find the asymptotic curves, i.e. the invariant manifolds from the X-point.

\begin{figure}[ht!]\centering
\includegraphics[scale=0.22]{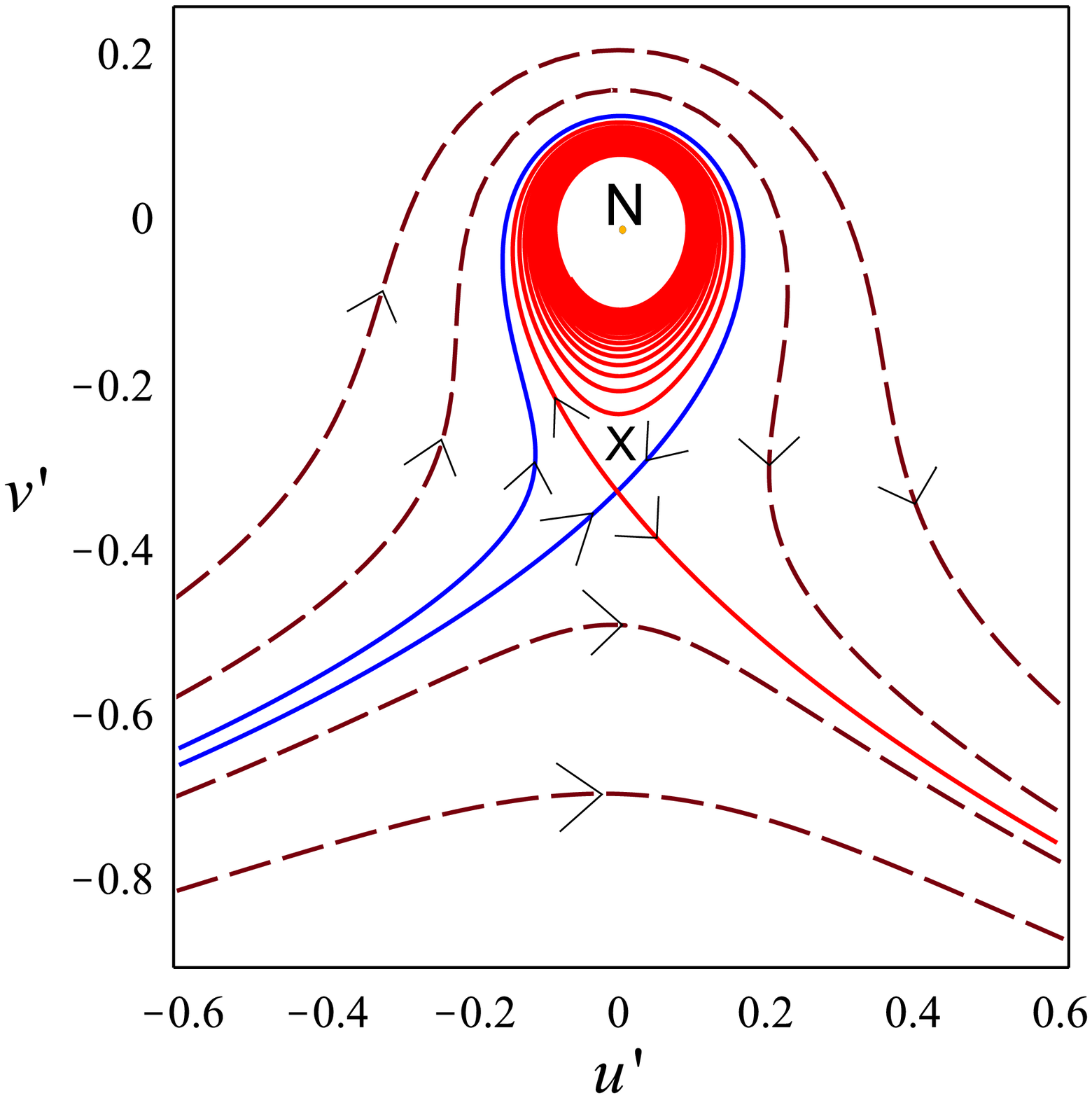}
\vspace*{-15mm}\caption{\label{fig:x_point_branches} The stable and unstable asymptotic curves from the X-point (blue and red respectively) for $t=4$ and $x_3'=4.23$ (continuous lines). One unstable asympotic curve spirals around the nodal point and tends to it asymptotically. More trajectories are shown (for other initial conditions) by dashed lines.}
\end{figure}

Figure \ref{fig:x_point_branches} shows an example of nodal point-X-point complex in our model for $t=4$. The dashed curves show more trajectories of (\ref{fic1}) besides the asymptotic curves (red and blue ones). Most of them pass above or below the nodal point X-point complex. Only a few trajectories (those starting between the two blue lines on the left) enter in the region between the nodal point and the X-point.
However, the form of the nodal point-Xpoint complex changes  with time. In particular, for some times the asymptotic curves that approach the nodal point are stable, while for other times they are unstable.  A transition from the first to the second case is shown in Figs~\ref{fig:hopf1}-{\ref{fig:hopf3}. 
This behavior is characteristic of the local form of the quantum flow after a `Hopf bifurcation' as discussed in figure (4) of Ref. \cite{PhysRevE.79.036203}.

\begin{figure}\label{fig:hopfg}
\captionsetup[subfloat]{captionskip=-2.5cm}
\subfloat[\label{fig:hopf1}]
{\includegraphics[height=3.2cm, width=2.95cm]{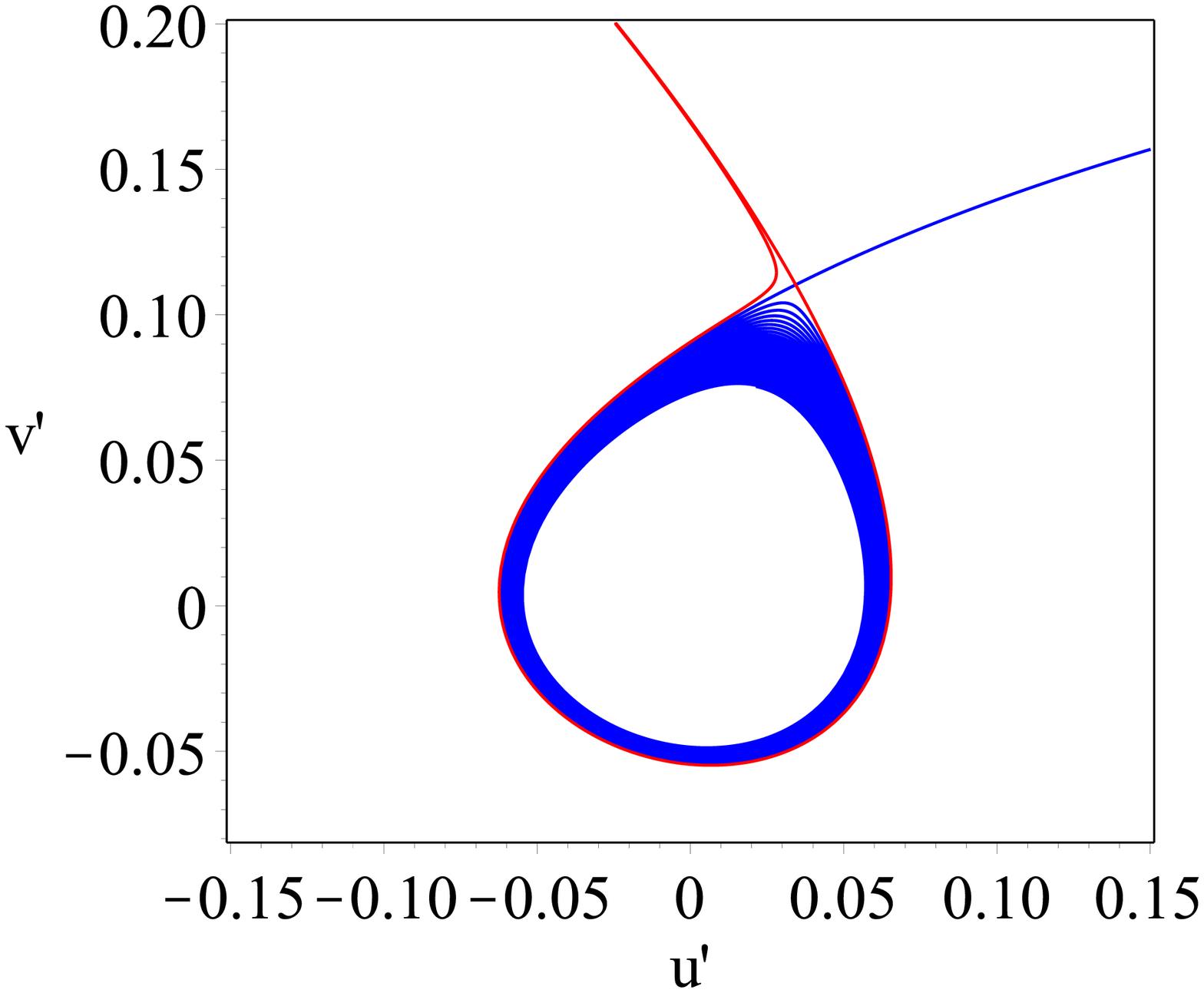}}\hfill
\subfloat[\label{fig:hopf2}]
{\includegraphics[height=3.2cm, width=2.95cm]{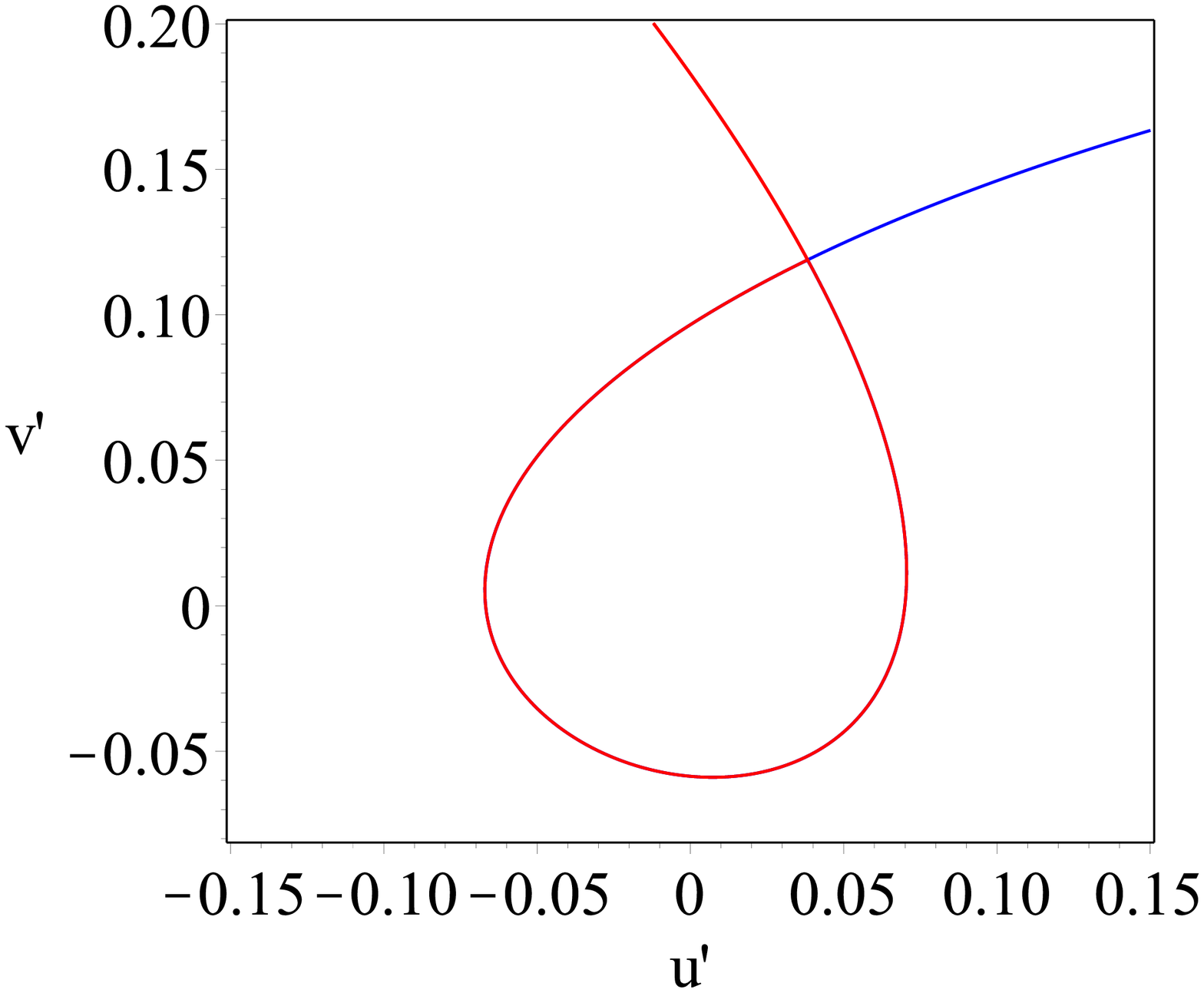}}
\subfloat[\label{fig:hopf3}]
{\includegraphics[height=3.2cm, width=2.95cm]{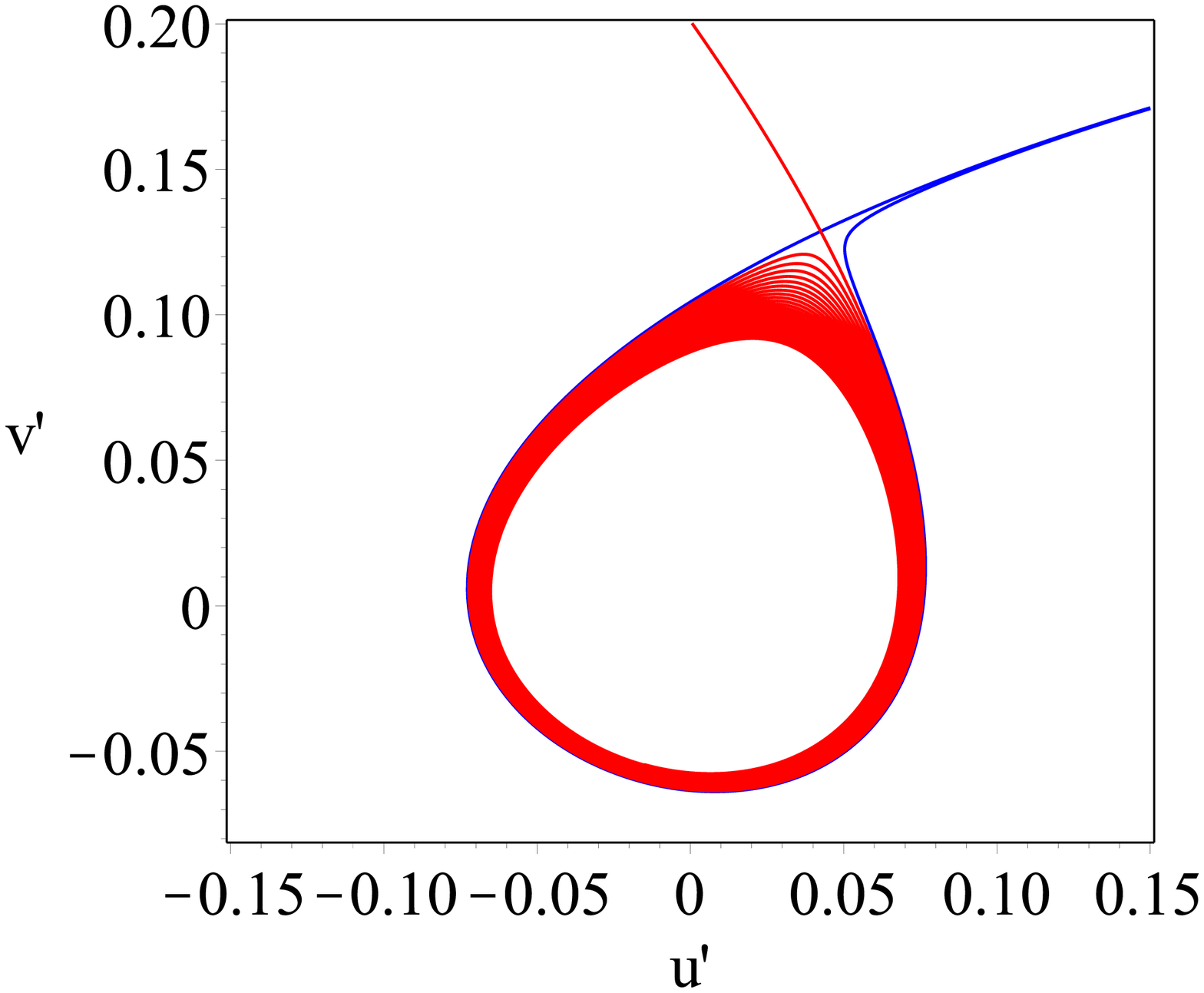}}
\caption{Time evolution of the nodal point-X-point complex around the nodal point at $(x_3'=5)$, in a case where the manifold approaching the nodal point from the corresponding X-point  turns from stable at $t=9.52$ (a) to  unstable at $t=9.6$ (c). The transition is made at the time $t=9.5586$ when the two manifolds coincide around the nodal point (b). }
\end{figure}

\textit{Trajectories:}
We now discuss the behavior of the trajectories near and far from a nodal point-X-point complex.

Figure~\ref{fig:troxia_1ews_100} shows a trajectory (blue) starting at a point close to the nodal point  of Fig. \ref{fig:troxia_nodal_point} and integrated numerically with the exact equations of motion (\ref{troxia1}) for $t=[1,100]$. The trajectory drifts on the sphere $R=4.23$, while it forms several loops on this surface. The moving nodal point on the same surface coincides  initially with the guiding center of this motion and remains so up to a time $t\simeq 7.4$. Beyond that time the form of the trajectory changes drastically. The trajectory forms now a box-like motion on the surface of the sphere. Comparing with Fig.\ref{fig:acceleration}, we see that the trajectory abandons the neighbourhood of the nodal point when the nodal point accelerates significantly (Fig.~\ref{fig:acceleration}). On the other hand, orbits starting far from the nodal point X-point complex form boxes on the surface of a sphere without exhibiting chaotic behavior (Fig.~\ref{fig:troxia_1ews_100_ordered}).

\begin{figure}[t!]\centering
\captionsetup[subfloat]{captionskip=-2.4cm}
\subfloat[\label{fig:troxia_1ews_100}]
{\includegraphics[scale=0.33]{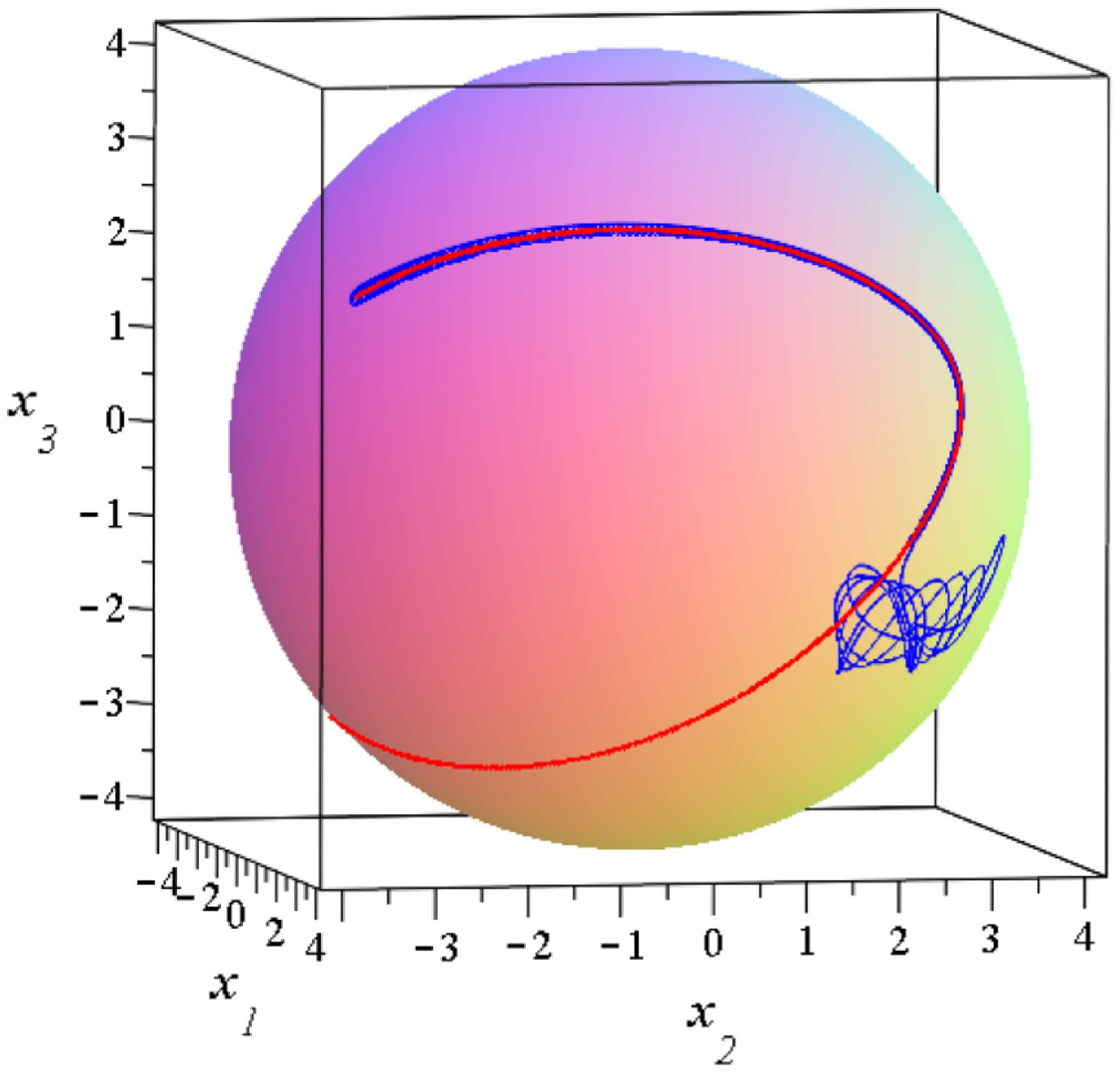}}
\subfloat[\label{fig:troxia_1ews_100_ordered}]
{\includegraphics[scale=0.33]{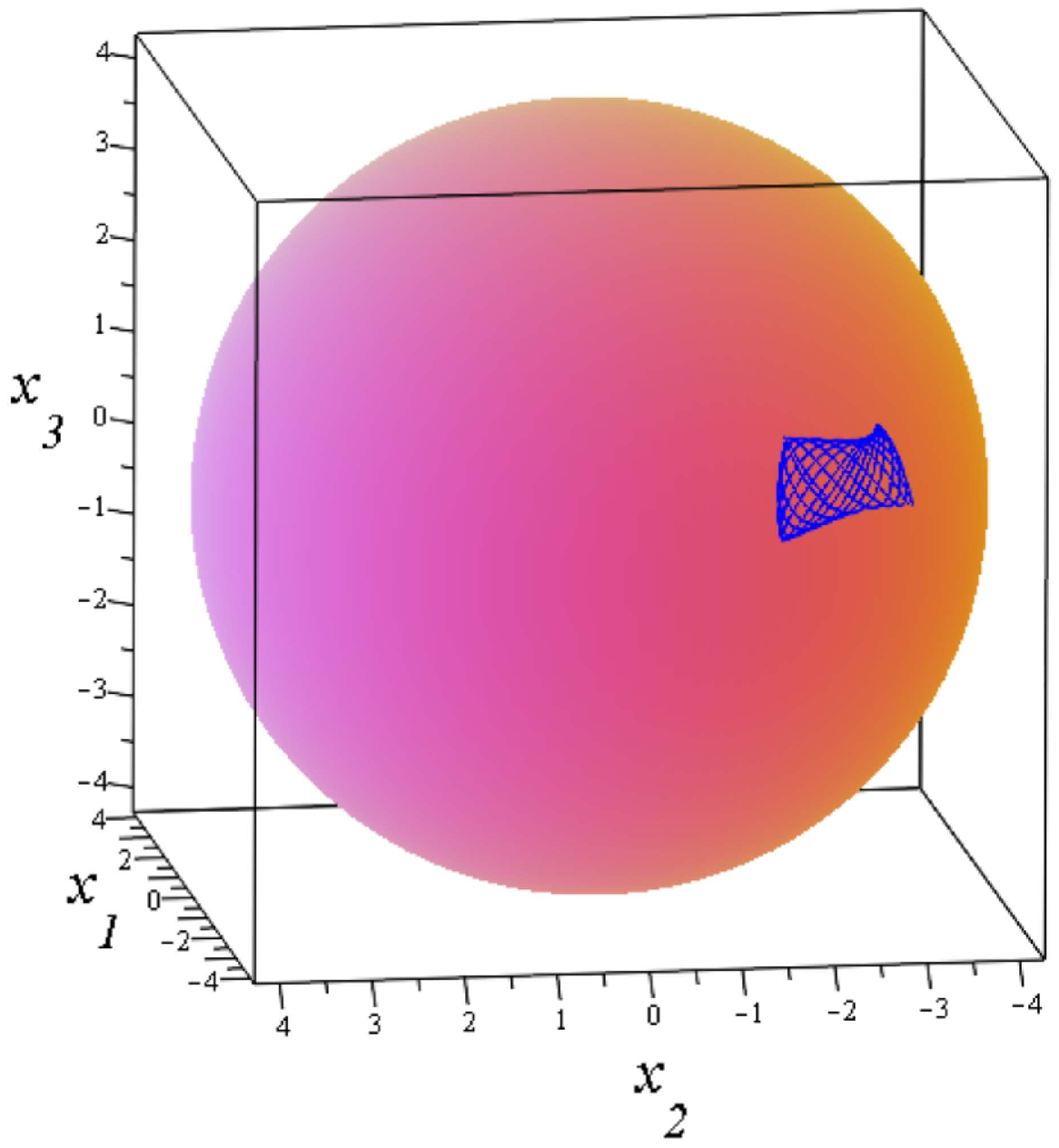}}
\caption{(a) A chaotic Bohmian trajectory starting at a distance $4.23$ from ($0,0,0,$) for $t=1\dots 100$. The trajectory forms a ribbon (blue) around the trajectory of the nodal point (red)  up to $t=7.4$. Then the trajectory deviates from the one of the nodal point and is trapped in a certain region of the same sphere. (b) A trajectory starting  away from the nodal lines for $t=1\dots 100$. This trajectory is ordered.}
\end{figure}

Figure~\ref{fig:orbits} shows five trajectories that start at different distances $R$ from the origin but at the same distance from the corresponding nodal point. All these orbits form loops  around the respective nodal points up to a certain time. In fact, as the distance from the origin increases, the  time needed for the trajectory to start derailing from the motion of the nodal line decreases. Similarly, Fig.~\ref{fig:orbit3} shows three trajectories that start close to the same nodal point  but at different distances from it. Then, for a smaller distance from the nodal point we get a longer time for the trajectory to get derailed. 

\begin{figure}[t!]\centering
\captionsetup[subfloat]{captionskip=-3.8cm,position=b}
\subfloat[]{\label{fig:orbits}\includegraphics[scale=0.14]{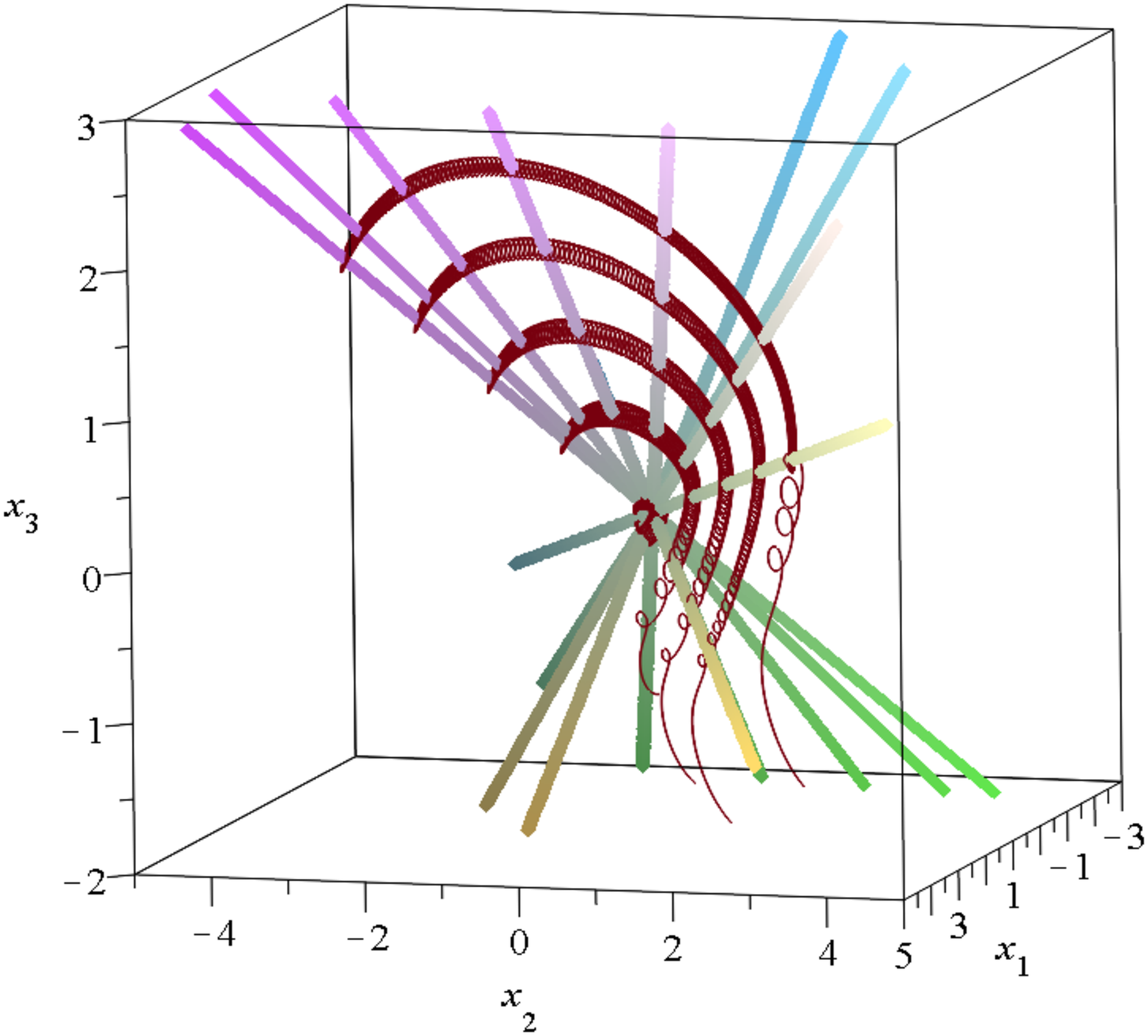}}
\subfloat[]{\label{fig:orbit3}\includegraphics[scale=0.15]{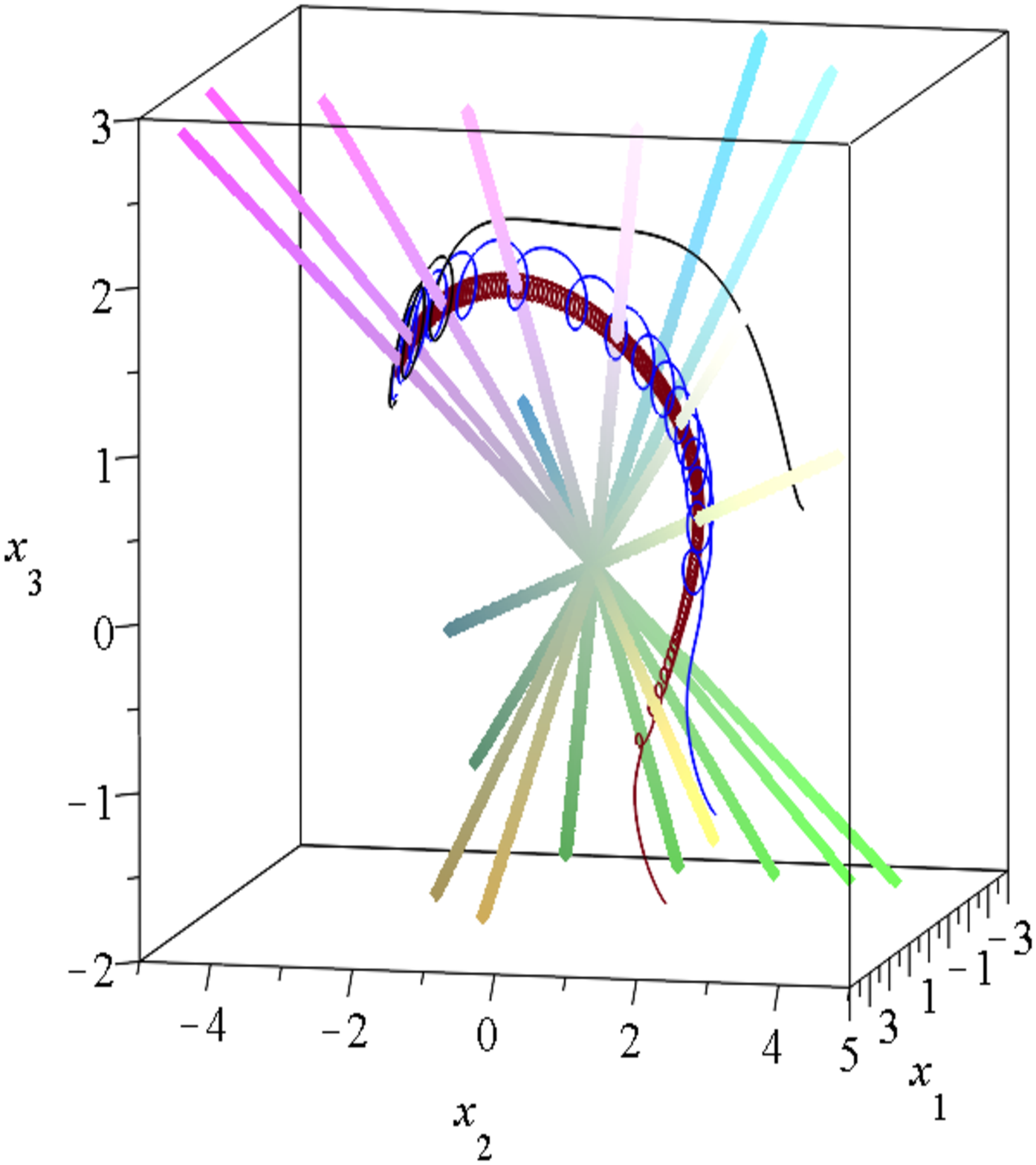}}
\caption{(a)  Bohmian trajectories for $t\in[1,10]$, which start close (at distance $d=0.1$) to various nodal points at $t=1$, for different distances $R$ from the origin, namely $R=0.2, 1.2, 2.2, 3.2$ and $4.2$. (b) Bohmian trajectories  starting close to the same nodal point but at different distances from it. Namely  $d=0.1$ (crimson), $d=0.25$ (blue) and  $d=0.3$ (black). Loops around the nodal point disappear faster as $d$ increases.}
\end{figure}

\textit{3d nodal point-X-point structures and the emergence of chaos:}
The nodal point-X-point complexes in our model are  2-d structures of the quantum flow, parameterized by the radius $R$ of the sphere upon which the considered nodal point moves. However, the foliation of all the planes orthogonal to a nodal line, for different $R$, forms a 3-d  structure containing the union of all nodal point-X-point complexes. This, we hereafter call the `3-d structure of nodal and X-points'.
An example of such a 3-d structure is shown in Fig.~\ref{fig:epipeda_xronos_4}. Along the structure, the complexes are aligned to each other in parallel layers.

\begin{figure}[t!]\centering
\includegraphics[scale=0.3]{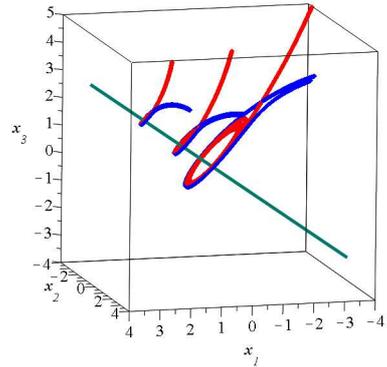}
\caption{\label{fig:epipeda_xronos_4} The foliation of nodal point -X-point complexes at $t=4$ forms a cylindrical structure around the nodal line (green).}
\end{figure}

The chaotic behavior of  trajectories as those given above can now be shown to be due to their scattering by the 3-d structure of nodal and X-points.
As an indicator of chaos, we use the Lyapunov Characteristic Number (LCN).
Let $\xi_k$ be the length of the deviation vector between two nearby trajectories at the time $t=k\tau,\,\,k=1,2,\dots$  The stretching number \cite{Contopoulos200210} is defined as
\begin{align}a_k=\ln\Big(\frac{\xi_{k+1}}{\xi_{k}}\Big)
\end{align}
and the finite time LCN is given by the equation: 
\begin{align}
\chi=\frac{1}{k\tau}\sum_{i=1}^k\ln a_i.
\end{align}
The LCN is the limit of $\chi$ when $k\to\infty$.
 Figure~\ref{fig:stretch1} shows an example of a  trajectory with an abrupt jump of the stretching number at $t\simeq 2$ (red curve). The black curve represents the cumulative stretching number $a_{cum}$ for the same time length, which shows the effect of the jump on the stretching number in the long run. The cumulative value after the end of the event ($t=6$) is stabilized to a positive value.  Furthermore, the distance between the X-point  and the trajectory becomes minimum at the time of the jump $t=t_{jump}\simeq 2$. Several such jumps are observed later, and the trajectory eventually stabilizes at a positive value of the LCN, i.e. becomes chaotic. Thus, the scattering by X-points is responsible for chaos in these trajectories.

\begin{figure}[t!]\centering
\includegraphics[scale=0.3]{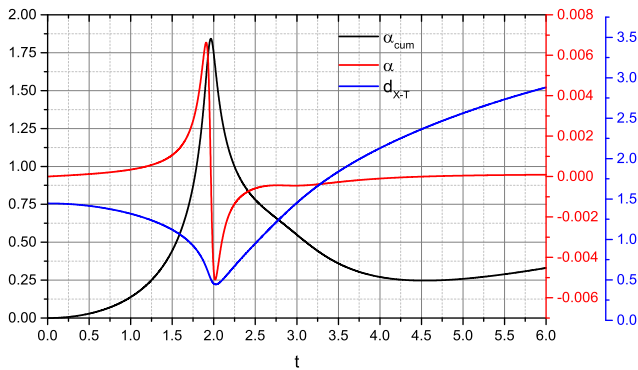}
\caption{\label{fig:stretch1} 
Local chaotic scattering of a trajectory by an X-point (initial conditions $x_1(0)=-1.5, x_2(0)=2, x_3(0)=-2$). We also compute  its deviation vector by numerically solving the variational equations (initial conditions $\delta x_1(0)=0, \delta x_2(0)=0, \delta x_3(0)=1$). The three curves show the value of the stretching number (red), cumulative stretching number (blue) and distance from the X-point (black) as functions of time. All three curves show a sudden jump at $t\simeq 2$, which marks the scattering event. The stretching number is computed with $\tau=1/100$.}
\end{figure}

\section{Chaotic 3d Diffusion}
Our results so far are based on a choice of  wavefunction which simplifies both the 3d structure of the nodal point-X-point complexes
and the shape of the Bohmian trajectories. Such a choice can serve as the leading approximation for more complex cases, where a small extra term added in the wavefunction leads to a disturbed form of the nodal point-X-point complexes as well as their resulting `3d structure of nodal and X-points'. 

As an example,  consider the perturbed wavefunction
\begin{align}\label{perturbed}
\Psi=& a_1\Psi_{1,0,0}+a_2\Psi_{0,1,0}+a_3\Psi_{0,0,1}+a_4\Psi_{0,0,2}
\end{align}
where $a_4$ is chosen small. We hereafter set $a_1=1/\sqrt{3}, a_2=1/\sqrt{3}$ and $ a_3=\sqrt{1-|a_1|^2-|a_2|^2-|a_4|^2}$.

An exact computation of the 3-d structure of nodal and X-points using this specific wavefunction (\ref{perturbed}) is computationally very difficult\footnote{Besides the need to find numerically the roots of Eqs.~(\ref{nodalsystem}) in every time step, in order to compute the X-points, one needs to compute also numerically the foliation of the locally orthogonal surfaces of \cite{falsaperla2003motion} along the whole  nodal line at every time.}. Nevertheless, we can assume that for $a_4$ small, this structure will not be very different from that of the unperturbed case (Fig.~\ref{fig:epipeda_xronos_4}). In Fig. \ref{fig:diataraxi_3d} the latter structure is plotted in the background, along with a chaotic trajectory of (\ref{perturbed}) for $a_4=0.05$. For the perturbed trajectory, the chaotic scattering effects still occur near the X-points. In particular, the trajectory (green curve, for $t\in[0,100]$) wanders chaotically in the configuration space, its scattering with the `3-d structure of nodal and X-points' taking now place at different distances $R$ from the origin.

The diffusion can be depicted by showing the time variation  of the distance $R$, which now exhibits chaotic variations due to the perturbation $a_4\neq 0$ (Fig. \ref{fig:diataraxi}). Careful inspection reveals that the chaotic variations are correlated with scattering events due to close approach to various X-points of the 3-d structure. In particular, the local jump of the stretching number (Fig. \ref{fig:stretch_perturbed}) at a scattering event has a profile quite similar to the one of the unperturbed case (red curve in Fig. \ref{fig:stretch1}). Furthermore, we can probe numerically the speed  of the chaotic diffusion  as a function of the perturbation $a_4$. Fig. \ref{fig:Delta_R_max} shows how the maximum jump $\Delta R_{max}$ of the variation curve of $R$ (as in Fig. \ref{fig:diataraxi}) varies with  $a_4$. We find an approximate power law  $\Delta R_{max}\sim a_4^p$ in the range $0<a_4<0.2$ with $p\simeq 1$.

\begin{figure}[t!]\centering
\subfloat[]{\label{fig:diataraxi_3d}\includegraphics[scale=0.16]{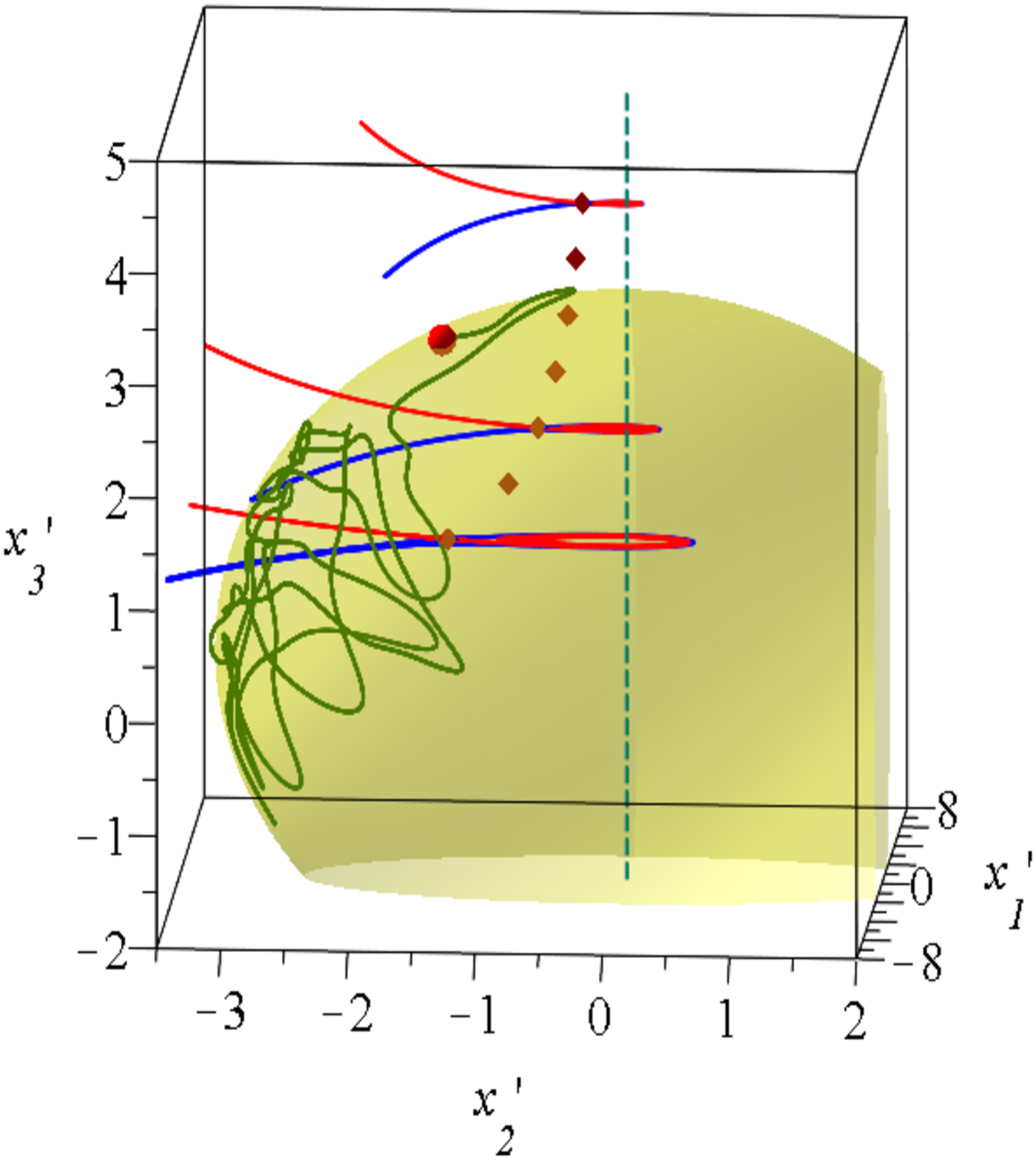}}
\subfloat[]{\label{fig:diataraxi}\includegraphics[height=4.7cm,width=3.6cm]{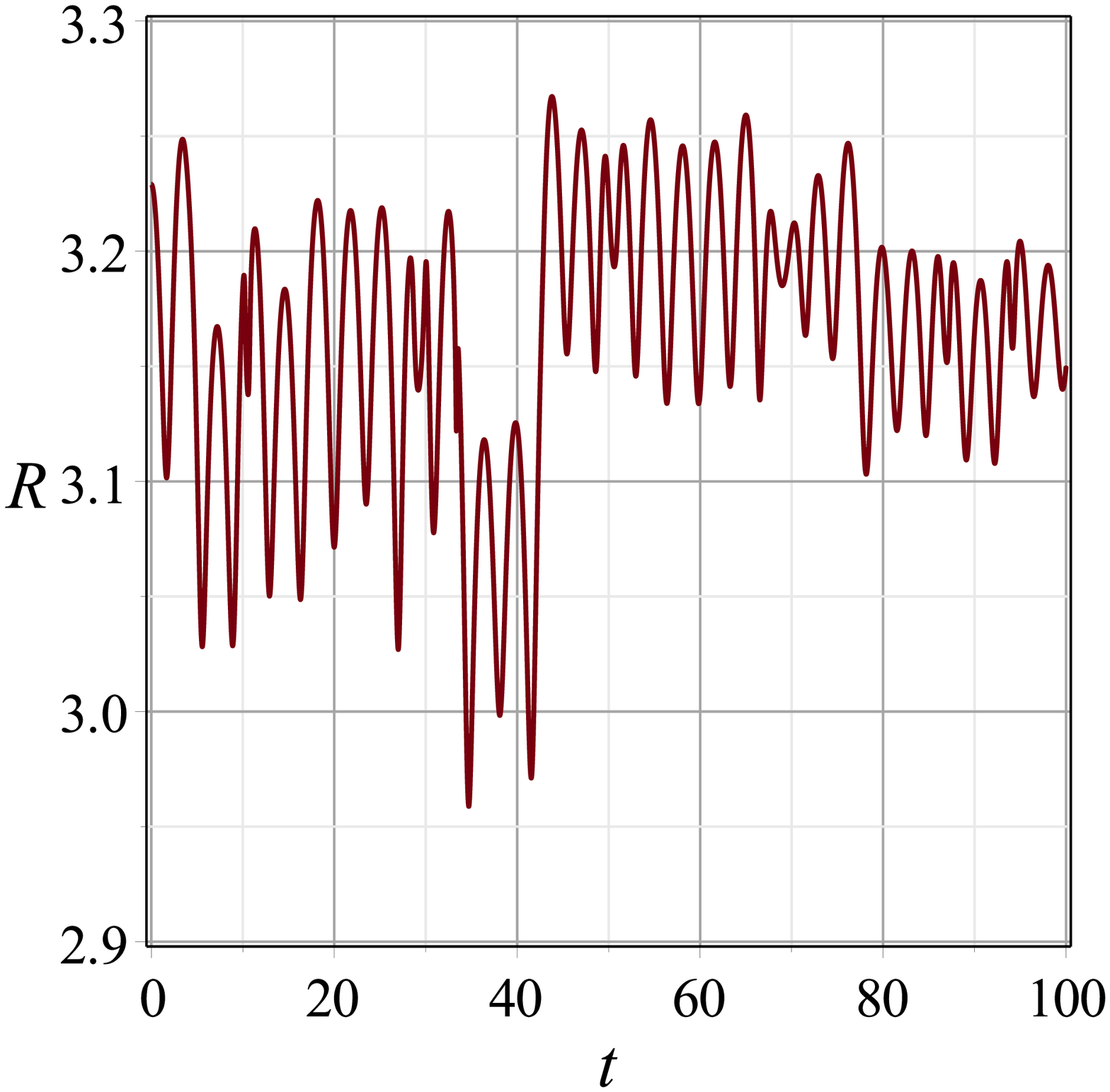}}\\
\subfloat[]{\label{fig:stretch_perturbed}\includegraphics[scale=0.2]{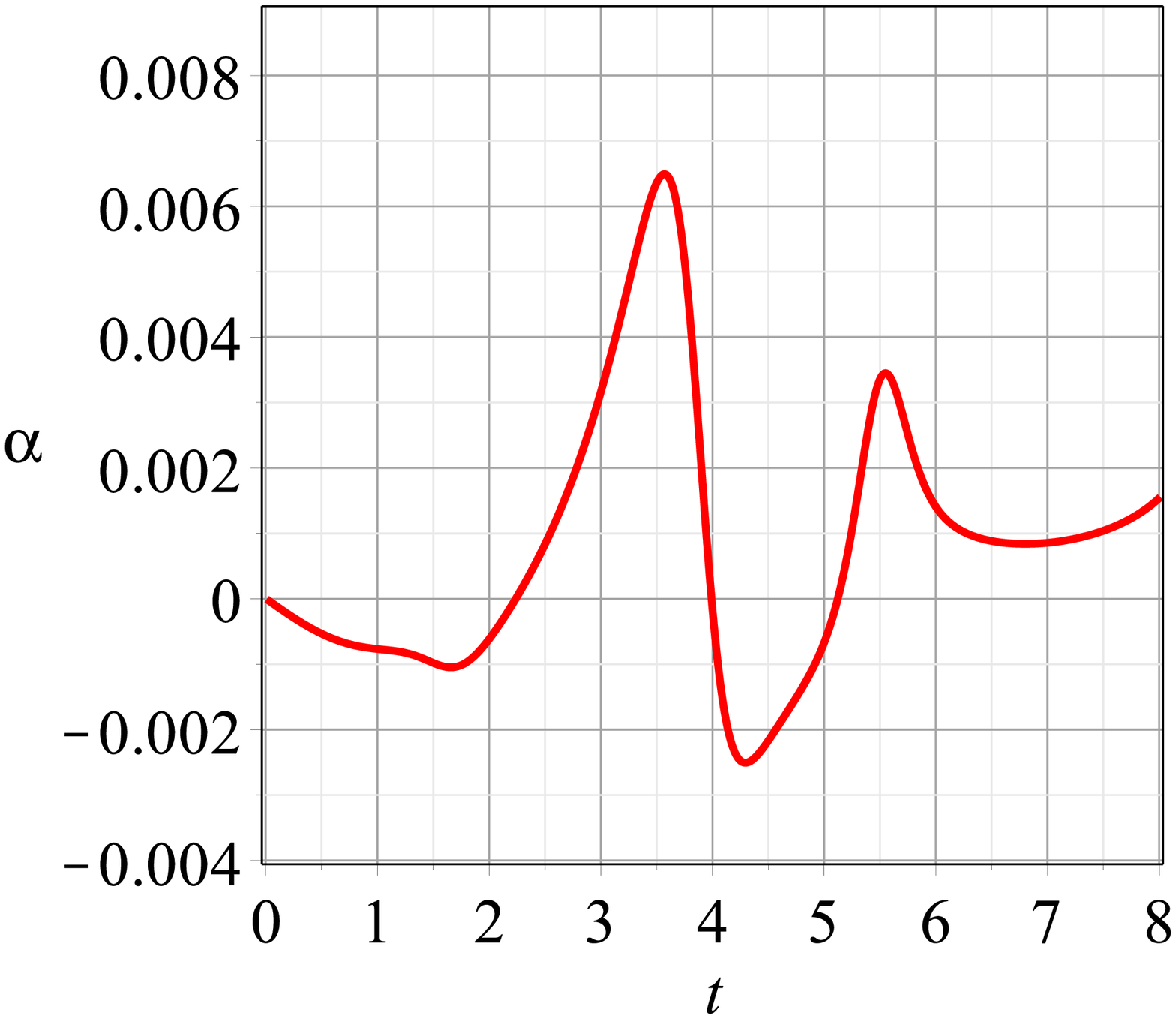}}\hfill
\subfloat[]{\label{fig:Delta_R_max} \includegraphics[scale=0.19]{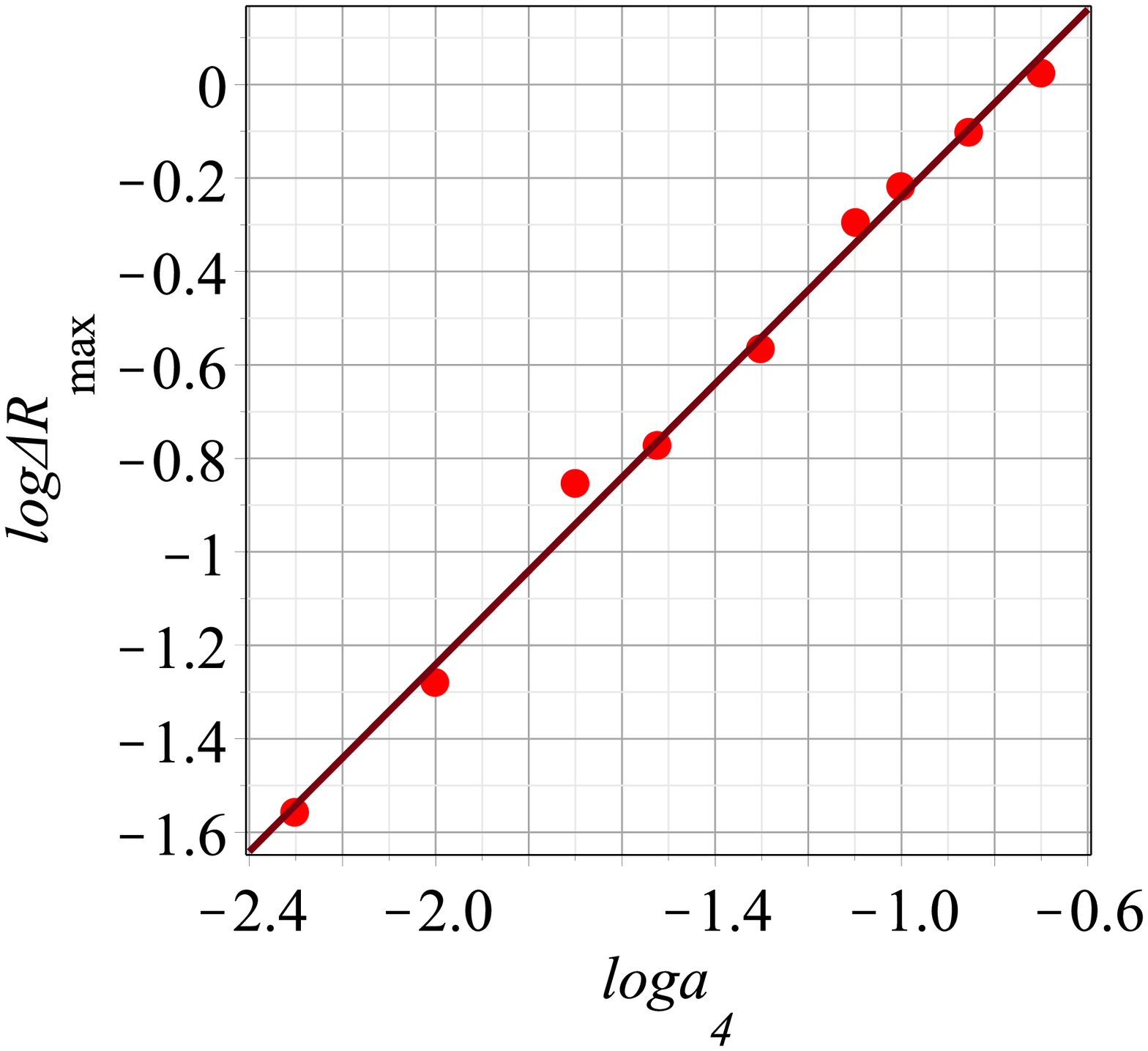}}
\caption{(a) A chaotic Bohmian trajectory (green curve) in the model (\ref{perturbed}) with $a=0.05$ and initial conditions $x_1(4)=2.2194, x_2(4)=-.4062, x_3(4)=2.3109$. The trajectory is shown in the coordinates $(x_1',x_2',x_3')$ of Eqs.~(\ref{transf}).  The dashed green line shows the nodal line and the three red-blue curves indicate the manifolds illustrating the corresponding nodal point-X-point structure. The squares indicate the X-line. (b) Variation of the distance $R(t)$ from the origin for the trajectory of (a). (c) A local scattering event (stretching number $a(t)$) around the value $t=4$ for the same trajectory.  The stretching number is computed with $\tau=1/100$. (d) Dependence of the maximum jump $\Delta R_{max}$ in the interval $0\leq t\leq 100$ on the perturbation $a_4$.}
\end{figure}

\section{Conclusions}
In this letter we propose a mechanism responsible for the generation of
chaos in the Bohmian trajectories of 3-d quantum systems. Our main results
are the following:

1. We give numerical (non-schematic) examples of the 3-d structure of
the quantum flow close to nodal points delineated along nodal curves,
as discussed already theoretically in \cite{falsaperla2003motion}. To this end, we employ 
a simple 3-d quantum model described by Eqs.(\ref{ham}-\ref{state}). When we pass
to a moving frame of reference attached to each nodal point, we find
that the quantum flow in a surface locally orthogonal to the nodal
curve contains a second critical point, whose character is always
hyperbolic, thus forming locally a `nodal point
- X-point complex'. Such a complex induces similar topological properties
of the quantum flow as those established theoretically in \cite{PhysRevE.79.036203} for
generic 2-d quantum systems. Also, the union of all the nodal points,
along with the X-points and the latter's stable and unstable manifolds,
forms a foliation in 3-d space, here called the `3-d structure of nodal
and X-points'. This structure complements in an essential way the
`cylindrical structure' of \cite{falsaperla2003motion}, by adding the X-points and their
manifolds to it. Here we only give numerical examples of it using a special model. However, we propose the question of how generic this structure is for future study.

2. We give several examples of numerically computed Bohmian trajectories,
both ordered and chaotic. The emergence of chaos is connected with the interaction of
the trajectories with the 3-d structure of nodal and X-points. In
particular, the rise of the Lyapunov exponent is caused by consecutive
`scattering events', taking place whenever a trajectory approaches
close to an X-point of the structure. Thus, contrary to the conjecture
of \cite{falsaperla2003motion}, chaos is possible even when only one such structure is present in the model.

3. Genuine 3-d chaotic diffusion occurs in models not subject to constraints of the Bohmian trajectories on invariant surfaces (as in Eq.~(\ref{sphere1})).
 In general, the approach of a trajectory
close to the X-line induces a local drift in the direction
{\it along} the 3-d structure of nodal and X-points. In the example of the model (\ref{perturbed})
the speed of the drift increases with the parameter $a_4$, which acts as a perturbation with respect to the model (\ref{state}) in which no such drift is possible.

{\bf Acknowledgements}: This research is supported by the Research Committee of the Academy of Athens.
\section*{References}
\bibliographystyle{elsarticle-num}
\bibliography{bibliography}

\end{document}